\documentclass{aa}  

\usepackage{graphicx}
\usepackage[dvipsnames]{xcolor}
\usepackage{caption}
\usepackage{multirow}
\usepackage{comment}
\usepackage{txfonts}
\usepackage[colorlinks,allcolors=blue]{hyperref}

\newcommand\figsize{0.48}

\newcommand{\xmm}{{\it XMM-Newton}\xspace}
\newcommand{\nustar}{{\it NuSTAR}\xspace}
\newcommand{\chandra}{{\it Chandra}\xspace}
\newcommand{\suzaku}{{\it Suzaku}\xspace}
\newcommand{\swift}{{\it Swift}\xspace}
\newcommand{\gaia}{{\it Gaia}\xspace}

\def\amin{\ifmmode^{\prime}\else$^{\prime}$\fi}
\def\asec{\ifmmode^{\prime\prime}\else$^{\prime\prime}$\fi}
\def\simgt{\lower.5ex\hbox{$\; \buildrel > \over \sim \;$}}
\def\simlt{\lower.5ex\hbox{$\; \buildrel < \over \sim \;$}}

\usepackage{threeparttable}

\usepackage{txfonts}
\begin{document}

   \title{\xmm and \nustar discovery of a likely IP candidate XMMU J173029.8--330920 in the Galactic Disk}

    \author{Samaresh Mondal\inst{1},
          Gabriele Ponti\inst{1,2},
          Luke Filor\inst{3},
          Tong Bao\inst{4},
          Frank Haberl\inst{2}, 
          Ciro Salcedo\inst{3},
          Sergio Campana\inst{1},
          Charles J. Hailey\inst{3},
          Kaya Mori\inst{3}, and 
          Nanda Rea\inst{5,6}
          }

   \institute{$^1$INAF -- Osservatorio Astronomico di Brera, Via E. Bianchi 46, 23807 Merate (LC), Italy
             \email{samaresh.mondal@inaf.it}\\
             $^2$Max-Planck-Institut f{\"u}r extraterrestrische Physik, Gie\ss enbachstra\ss e 1, 85748, Garching, Germany\\
             $^3$Columbia Astrophysics Laboratory, Columbia University, New York, NY 10027, USA\\
             $^4$School of Astronomy and Space Science, Nanjing University, Nanjing 210046, China\\
             $^5$Institute of Space Sciences (ICE, CSIC), Campus UAB, Carrer de Can Magrans s/n, E-08193 Barcelona, Spain\\
             $^6$Institut d'Estudis Espacials de Catalunya (IEEC), Carrer Gran Capit\`a 2--4, 08034 Barcelona, Spain
             }

   \date{Received XXX; accepted YYY}
   \authorrunning{Mondal et al.}
   \titlerunning{Discovery of the new IP XMMU J173029.8--33092}

 
  \abstract
   {}
   {We aim at characterizing the population of low-luminosity X-ray sources in the Galactic plane by studying their X-ray spectra and periodic signals in the light curves.}
   {We are performing an X-ray survey of the Galactic disk using \xmm, and the source XMMU J173029.8--330920 was serendipitously discovered in our campaign. We performed a follow-up observation of the source using our pre-approved \nustar target of opportunity time. We used various phenomenological models in {\sc xspec} for the X-ray spectral modeling. We also computed the Lomb-Scargle periodogram to search for X-ray periodicity. A Monte Carlo method was used to simulate 1000 artificial light curves to estimate the significance of the detected period. We also searched for X-ray, optical, and infrared counterparts of the source in various catalogs.}
   {The spectral modeling indicates the presence of an intervening cloud with $N_{\rm H}\sim(1.5-2.3)\times10^{23}\ \rm cm^{-2}$ that partially absorbs the incoming X-ray photons. The X-ray spectra are best fit by a model representing emission from a collisionally ionized diffuse gas with plasma temperature $kT=26^{+11}_{-5}$ keV. Furthermore, an Fe $K_{\alpha}$ line at $6.47^{+0.13}_{-0.06}$ keV was detected with an equivalent width of the line of $312\pm104$ eV. We discovered a coherent pulsation with a period of $521.7\pm0.8$ s. The 3--10 keV pulsed fraction of the source is around $\sim$50--60\%.}
   {The hard X-ray emission with plasma temperature $kT=26^{+11}_{-5}$ keV, iron $K_{\alpha}$ emission at 6.4 keV and a periodic behavior of $521.7\pm0.8$ s suggest XMMU J173029.8--33092 to be an intermediate polar. We estimated the mass of the central white dwarf to be $0.94-1.4\ M_{\odot}$ by assuming a distance to the source of $\sim1.4-5$ kpc.}


   \maketitle
   
%
\section{Introduction}
Galactic X-ray emission is a manifestation of various high-energy phenomena and processes. Studying discrete X-ray sources is essential to understand stellar evolution, dynamics, end-products, and accretion physics. Initial X-ray scans of the Galactic plane revealed a narrow, continuous ridge of emission, the so-called Galactic ridge X-ray emission (GRXE; \citealt{worrall1982,warwick1985}), which extends either side of the Galactic disk up to $\sim40^\circ$. A copious amount of 6.7 keV line emission from highly ionized iron was detected from selected regions of the GRXE \citep{koyama1989,yamauchi1990}. The study of the 6.7 keV emission indicated an optically thin hot plasma distributed mainly along the Galactic plane. The Galactic diffuse X-ray emission can be described by a two-component emission model with a soft temperature of $\sim0.8$ keV and a hard component of $\sim8$ keV \citep{komaya1996,tanaka2000}. The origin of the $\sim8$ keV plasma is less certain. The Galactic potential is too shallow to confine such hot plasma that would escape at a velocity of thousand km s$^{-1}$. The Deep \chandra observations of the Galactic Center (GC) by \citet{wang2002} ($2^\circ\times0.8^\circ$) and a region south of the GC ($16'\times16'$) by \citet{revnivtsev2009}, pointed out that the hard $kT\sim8$ keV diffuse emission is primarily due to unresolved cataclysmic variables (CVs) and active binaries. However, later, close to the GC within $\pm0.8^\circ$ a significant diffuse hard X-ray emission was detected up to energies of 40 keV \citep{yuasa2008}. In addition, some studies pointed out that the hard GC emission is truly diffuse by comparing stellar mass distribution with the Fe XXV (6.7 keV) line intensity map \citep{uchiyama2011,nishiyama2013,yausi2015}. A recent study by our group showed that this diffuse hard emission in the GC can be explained if one assumes the GC stellar population with iron abundances $\sim1.9$ times higher than those in the Galactic bar/bulge \citep{anastasopoulou2023}.

Magnetic CVs (mCVs) are categorized into two types based on the magnetic field strength of the white dwarf (WD): intermediate polars (IPs) and polars \citep[see][for reviews]{copper1990,petterson1994,mukai2017}. In polars, the magnetic field is strong ($>10$ MG) enough to synchronize the spin and orbital period of the system. Whereas in IPs, the magnetic field is less strong; therefore, their spin and the orbital period are less synchronized. 
Most hard X-ray emission from mCVs originates close to the WD surface. The accreting material from the companion star follows the magnetic field lines of the WD and, while approaching the WD polar cap, reaches a supersonic speed of 3000--10000 km $\rm s^{-1}$. A shock front is developed above the star, and the in-falling gas releases its energy in the shock, resulting in hard X-ray photons \citep{aizu1973}.

The Galactic Center (GC) region hosts many energetic events, like bubbles and super-bubbles from young and old stars and various supernovae remnants \citep{ponti2015} as well as large-scale structures like Galactic Chimneys \citep{ponti2019,ponti2021}, \emph{Fermi} bubbles \citep{su2010}, and eROSITA bubbles \citep{predehl2020}. Currently, we are performing an X-ray survey of the Galactic disk using \xmm. The main aim of this survey is to constrain the flow of hot baryons that feed the large-scale structures from the GC. The survey region extends from $l\geq350^\circ$ to $l\leq7^\circ$ and covers a latitude of $b\sim\pm1^\circ$. The survey has an exposure of 20 ks per tile. During this survey, we detected thousands of X-ray point sources of various types. Our survey is sensitive enough to detect sources as faint as  $10^{-14}\ \rm erg\ cm^{-2}\ s^{-1}$. The nearby bright X-ray point sources with flux $>10^{-11}\ \rm erg\ cm^{-2}\ s^{-1}$ are well-known thanks to several decades of monitoring by \emph{RXTE}, \emph{Swift}-BAT, \emph{MAXI}, and \emph{INTEGRAL}. The bright IPs in the solar neighborhood have a typical luminosity of $10^{33}-10^{34}\ \rm erg\ s^{-1}$ \citep{downes2001,ritter2003,suleimanov2022} and its unabsorbed flux at the GC would be $\sim10^{-13}-10^{-12}\ \rm erg\ cm^{-2}\ s^{-1}$. Furthermore, the high extinction towards the GC makes it even more difficult to detect these faint sources in the GC. Our \xmm observations are deep enough to detect these faint X-ray sources. During our \xmm campaign, we detected a faint X-ray source, XMMU J173029.8--33092, with a 3--10 keV X-ray luminosity of $\sim7\times10^{-13}\ \rm erg\ cm^{-2}\ s^{-1}$. In this paper, we report the detailed spectral and timing study of this source. 

\section{Observations and data reduction}
The source XMMU J173029.8--330920 ($\rm R.A.=17^{h}30^{m}29\fs29$, $\rm DEC=-33\degr09\arcmin20\farcs9$) was discovered in our \xmm \citep{jansen2001} \emph{Heritage} survey of the inner Galactic disk (PI: G. Ponti). The source was detected in one of our \xmm pointings on 2023-03-28 (ObsID: 0916800201) with an exposure of 18.3 ks \citep{mondal2024}. We analyzed the \xmm data and discovered an X-ray pulsation and a tentative iron 6.4 keV line emission that suggested a possible identification with an IP. To confirm this, we triggered our pre-approved 35 ks \nustar target of opportunity observation to follow up on this source. 

The \xmm observation data file was processed using \xmm Science Analysis System (SASv19.0.0\footnote{https://www.cosmos.esa.int/web/xmm-newton/sas}). The point source detection and source list creation were performed using the SAS task \texttt{emldetect}. The source detection was performed simultaneously in five energy bands (0.2--0.5 keV, 0.5--1 keV, 1--2 keV, 2--4 keV, and 4--12 keV) for each detector EPIC-pn/MOS1/MOS2 \citep{struder2001,turner2001}. While extracting the source products, we only selected events with PATTERN$\leq$4 and PATTERN$\leq$12 for EPIC-pn and MOS2 detectors, respectively. The source was out of the field of view of the MOS1 detector. We applied barycenter correction for light curve extraction using the SAS task \texttt{barycen}. We used a circular region of 25\arcsec\ radius for the source product extraction. The background products were extracted from a nearby source-free circular region of radius 25\arcsec.

The \nustar \citep{harrison2013} observation was performed on 2023-04-24 (ObsID: 80801332002, PI: S. Mondal). The \nustar data reduction was performed using the \nustar data analysis pipeline software (NUSTARDAS v0.4.9 provided with HEASOFT). The unfiltered event files from FMPA and FPMB detectors were cleaned, and data taken during passages through the South Atlantic Anomaly were removed using \texttt{nupipeline}. The source products were extracted from a circular region of radius 40\arcsec\ using \texttt{nuproducts}. The background was selected from a source-free region of 40\arcsec\ radius.
\begin{figure}
    \centering
    \includegraphics[width=\figsize\textwidth]{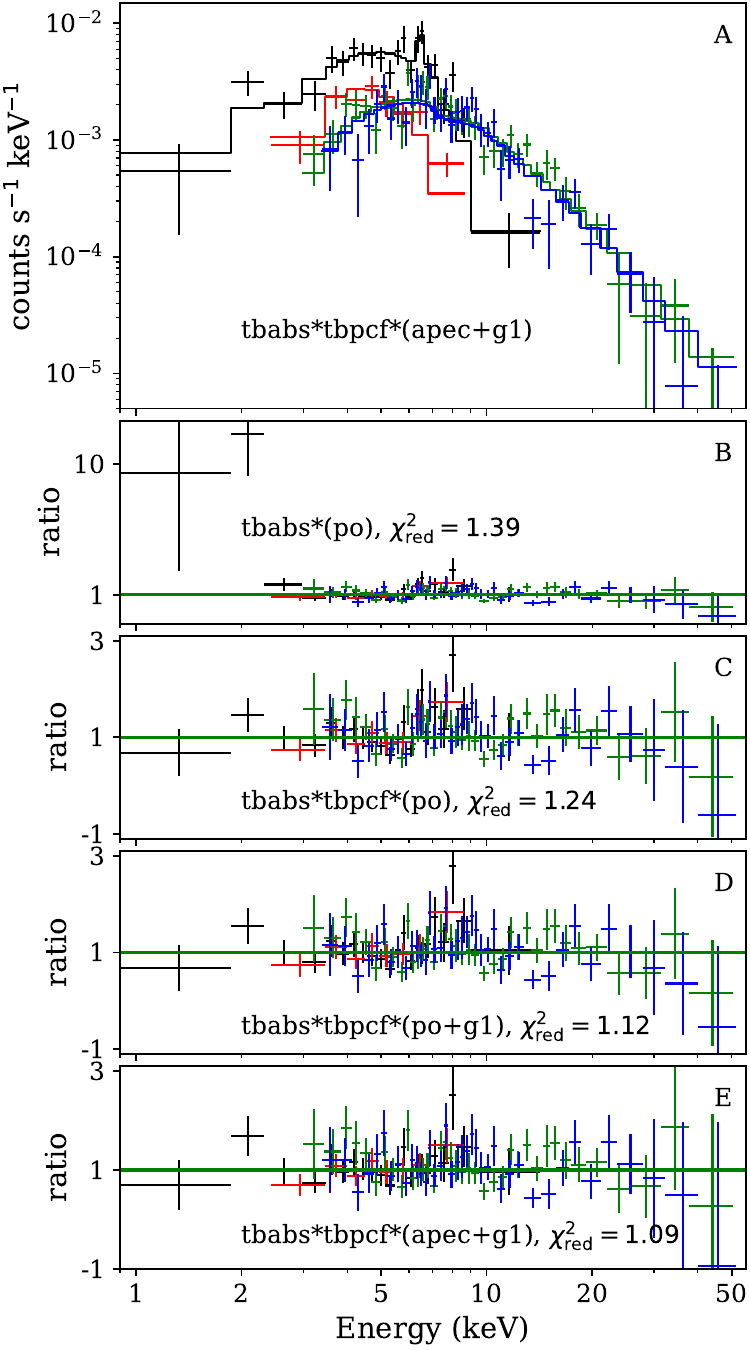}
    \caption{The spectra of XMMU J173029.8--330920, fitted with various partially absorbed models plus a Gaussian component at $\sim$6.4 keV. The black, red, green, and blue data points are from EPIC-pn, MOS2, FPMA, and FPMB detectors. \texttt{tbabs} and \texttt{tbpcf} represents the Galactic and partial absorption component. \texttt{po} (a simple power law) and \texttt{apec} (a collisionally ionized diffuse gas) are the two continuum models tested in the spectral modeling; more details are given in Sect. \ref{sepcmodel}. }
    \label{fig:ip2023_ratio}
\end{figure}
\section{Results}
\subsection{\xmm and \nustar joint spectral modeling}
\label{sepcmodel}
We performed a time-averaged spectral modeling using {\sc xspec} \citep{arnaud1996}, including the data from EPIC-pn, MOS2, FPMA, and FPMB detectors in a joint fit. While doing the fit, we added a constant term for cross-calibration uncertainty, which we fixed at unity for the EPIC-pn detector and free for others. The best-fit parameters are listed in Table \ref{table:fit_tab} with the quoted errors at the 90\% confidence level. All the models are convolved with a Galactic absorption component \texttt{tbabs} \citep{wilms2000} with solar abundance.

Before we started fitting the data from \xmm and \nustar jointly, we checked for any variation in flux, as the two observations are separated by nearly 27 days. We chose a common energy band of 3--10 keV in \xmm and \nustar to estimate the flux and fit the spectra using a simple power law model. The power law model provides 3--10 keV flux in \xmm and \nustar is $7.1^{+0.7}_{-1.7}\times10^{-13}\ \rm erg\ s^{-1}\ cm^{-2}$ and $7.4^{+0.6}_{-1.0}\times10^{-13}\ \rm erg\ s^{-1}\ cm^{-2}$, respectively. As we do not see any indication of flux variation, we jointly modeled the spectra from \xmm and \nustar. First, we fit the data with an absorbed power law model that leads to a high $N_{\rm H}=(16\pm3)\times10^{22}\ \rm cm^{-2}$ and $\Gamma=1.9\pm0.2$. Fitting the data with this model reveals residuals at energies below 3 keV. The partial covering component improves the fit by $\Delta\chi^2=17$ for two additional degrees of freedom ($d.o.f$). The significance of the partial covering component is 99.84\% in an F-test. The fitted parameters are $N_{\rm H}=3^{+2}_{-1}\times10^{22}\ \rm cm^{-2}$, $N_{\rm H,pcf}=23^{+8}_{-6}\times10^{22}\ \rm cm^{-2}$, $\rm pcf=0.93^{+0.04}_{-0.07}$ and $\Gamma=2.1\pm0.2$. Next, we add a Gaussian that further improves the fit by $\Delta\chi^2=13$ for an additional one $d.o.f$ that gives a detection significance of 99.90\% in an F-test. The line width is consistent with zero; therefore, we set the line width at zero keV, and the centroid of the line is estimated as $E_{\rm g}=6.47^{+0.13}_{-0.06}$ keV. The equivalent width of the line is $312\pm104$ eV. We checked the presence of Fe XXV and Fe XXVI by adding two Gaussians at 6.7 and 6.9 keV; however, there is no significant improvement in the fit. The best fit is obtained when the data is fitted with an absorbed \texttt{apec} model, representing the emission from collisionally-ionized diffuse gas. The \texttt{apec} model does not include the neutral Fe $K_{\alpha}$ emission at 6.4 keV. Therefore, we add a Gaussian at 6.4 keV when fitting with this model. The plasma temperature obtained from this model is $26^{+11}_{-5}$ keV.

\begin{table*}
\caption{The best-fit parameters of the fitted models.}
\label{table:fit_tab}
\setlength{\tabcolsep}{5pt}                   
\renewcommand{\arraystretch}{2.0}               
\centering
\begin{tabular}{c c c c c c c c c}
\hline\hline
 \multirow{2}{*}{\texttt{tbabs}*Model} & $N_{\rm H}$ & $N_{\rm H,pcf}$ & \multirow{2}{*}{$pcf$} & $\Gamma/kT$ & \multirow{2}{*}{$N$} & $E_{\rm g}$ & \multirow{2}{*}{$N_{\rm g}$} & \multirow{2}{*}{$\chi^2/d.o.f$}\\ 
 & $\times10^{22}$ & $\times10^{22}$ && -/keV && keV &&\\ \hline

 \texttt{po} & $16\pm3$ &&& $1.9\pm0.2$ & $4^{+2}_{-1}\times10^{-4}$ &&& 140/101\\ \hline
 \texttt{tbpcf*po} & $3^{+2}_{-1}$ & $23^{+8}_{-6}$ & $0.93^{+0.04}_{-0.07}$ & $2.1\pm0.2$ & $7^{+6}_{-3}\times10^{-4}$ &&& 123/99\\ \hline
 \texttt{tbpcf*(po+g1)} & $3^{+2}_{-1}$ & $19^{+7}_{-5}$ & $0.92^{+0.05}_{-0.08}$ & $1.9\pm0.2$ & $5^{+4}_{-2}\times10^{-4}$ & $6.47^{+0.13}_{-0.06}$ & $(3\pm1)\times10^{-6}$ & 110/98 \\ \hline
 \texttt{tbpcf*(apec+g1)} & $2^{+2}_{-1}$ & $15^{+5}_{-4}$ & $0.89^{+0.14}_{-0.07}$ & $26^{+11}_{-5}$ & $(8\pm1)\times10^{-4}$ & $6.45^{+0.09}_{-0.06}$ & $(4\pm2)\times10^{-6}$ & 106/97\\
 
\hline
\end{tabular}
\tablefoot{$N_{\rm H}$ is given in units of $10^{22}$ cm$^{-2}$, $N$ in photons keV$^{-1}$ cm$^{-2}$ s$^{-1}$ at 1 keV and $N_{\rm g}$ in photons keV$^{-1}$ cm$^{-2}$ s$^{-1}$ at the line energy. For the \texttt{apec} model, the metal abundance value is frozen to 1.0.}
\end{table*}

\subsection{X-ray pulsation search}
\begin{figure}
    \centering
    \includegraphics[width=\figsize\textwidth]{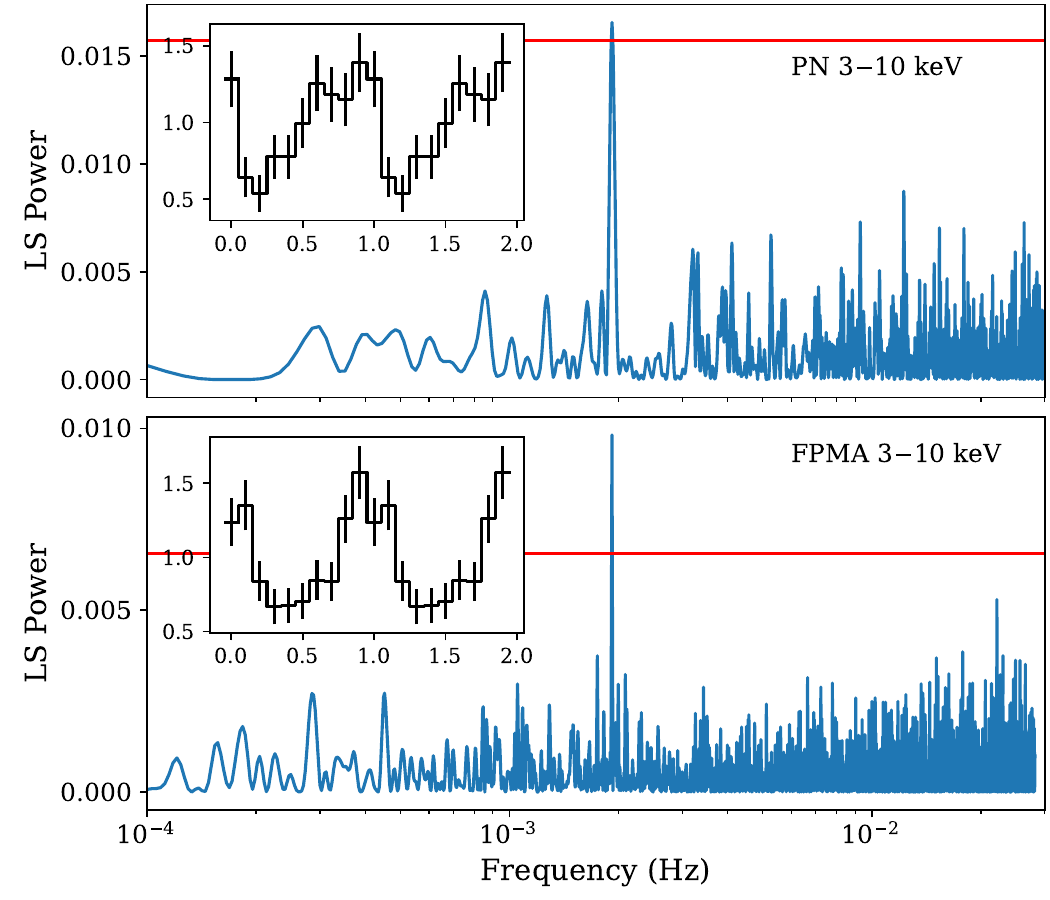}
    \caption{The Lomb-Scargle periodogram of the source XMMU J173029.8--330920. The top panel is for data from the \xmm EPIC-pn detector, and the bottom panel is for the \nustar FPMA detector. In both cases, a significant peak around $1.918\times10^{-3}$ Hz is visible, which corresponds to the period of 521 s. The small insects show the 3--10 keV folded light curve. The red horizontal lines indicate the $3\sigma$ detection significance.}
    \label{fig:ip2023_psd}
\end{figure}

For a primary analysis and search for X-ray pulsation in \xmm EPIC-pn and \nustar FPMA detectors, we chose a common energy band 3--10 keV to extract the light curve. The X-ray light curves often suffer from gaps due to the South Atlantic Anomaly Passage. The gaps in the light curve are prominent for satellites in low earth orbit such as \nustar due to Earth occultation. We used the Lomb-Scargle \citep{lomb1976} periodogram to search for periodicity. The Lomb-Scargle periodogram is well known for detecting periodicity in observations that suffer from gaps. Figure \ref{fig:ip2023_psd} shows the Lomb-Scargle periodogram for the EPIC-pn and FPMA detector. A peak at around a frequency of $1.918\times10^{-3}$ Hz and $1.916\times10^{-3}$ Hz is visible in EPIC-pn and FPMA detectors, respectively. The periods obtained from the EPIC-pn and FPMA data are $521.2\pm3.6$ and $521.7\pm0.8$, respectively; both are consistent with each other. 

The periodogram of accreting X-ray binaries displays red noise, which can mimic an artificial periodic or aperiodic variability. Therefore, we did Monte Carlo simulations to test the detection significance. To begin, we computed the power spectral density (PSD) using the standard periodogram approach with an $\rm [rms/mean]^2$ PSD normalization \citep{2003MNRAS.345.1271V}. Subsequently, we employed a power law model to characterize the source power spectrum, taking into account the presence of red noise. The power law model is expressed as:
\begin{equation}
P(\nu)=N \nu^{-1}+C_{\rm p},
\label{eqn:LMXBps}
\end{equation}
where $N$ represents the normalization factor, and $C_{\rm p}$ accounts for the Poisson noise, which is influenced by the mean photon flux of the source. We fitted the PSD of the source using Eq.~\ref{eqn:LMXBps} with a Markov Chain Monte Carlo approach, employing the Python package \emph{emcee} \citep{2013PASP..125..306F} to derive the best-fit parameters and their associated uncertainties. We simulated 1000 light curves for this best-ﬁt power law model using the method of \citet{1995A&A...300..707T}, which were re-sampled and binned to have the same duration, mean count rate, and variance as the observed light curve. The red horizontal lines in Fig. \ref{fig:ip2023_psd} indicate the $3\sigma$ detection significance.

\begin{figure}
    \centering
    \includegraphics[width=\figsize\textwidth]{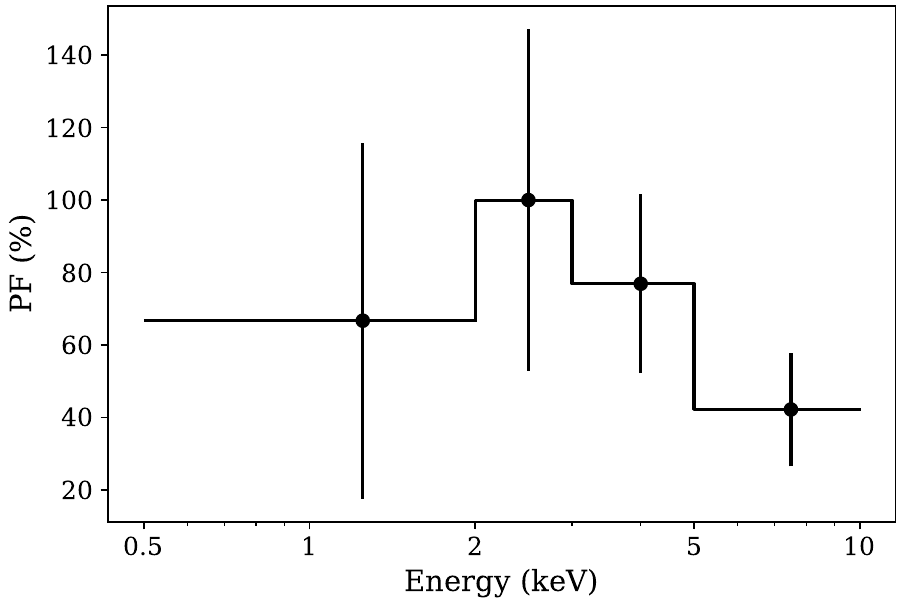}
    \caption{The PF of the source as a function of energy. The error bars of the pulse fraction are on 1$\sigma$ confidence level.}
    \label{fig:ip2023_pf}
\end{figure}

The 3--10 keV pulsed fractions (PF) obtained from EPIC-pn and FPMA are $50.4\pm15.5$ and $59.3\pm14.7$, respectively. The PF is calculated by using the formula $\rm PF=\frac{F_{max}-F_{min}}{F_{max}+F_{min}}$, where $\rm F_{max}$ and $\rm F_{min}$ are the normalized count of the folded light curve. Next, we computed the PF of the source at different energies. Figure \ref{fig:ip2023_pf} shows the pulsed fraction at different energies. The 0.5--5 keV and 5--10 keV pulse fraction is $55\pm17\%$ and $42\pm16\%$, respectively.

\section{Counterparts}
We searched for X-ray counterparts in \chandra CSC2.0\footnote{http://cda.cfa.harvard.edu/cscweb/index.do} and \swift 2SXPS\footnote{https://www.swift.ac.uk/2SXPS/index.php} catalogs, however, no counterparts were found. We searched for optical and infrared counterparts in various catalogs. No \gaia counterpart was found within 5\arcsec from \xmm position due to high ISM absorption. A source is found in the VVV Infrared Astrometric Catalogue \citep{smith2018}; however, the source is located 2.7\arcsec\ away from the \xmm position, which is off by more than $3\sigma$ (1\arcsec) \xmm positional uncertainty. So, we consider no infrared counterparts were detected. From the observed column density of absorption of $N_{\rm H}=2^{+2}_{-1}\times10^{22}\rm \ cm^{-2}$, we estimate the distance to the source by comparing this value to the absorption column density obtained from several other tracers \citep{planck2016,yao2017,edenhofer2023}. Assuming that all the observed column density in the X-ray band is due to interstellar absorption, we obtained the distance to the source is around $2.9^{+2.1}_{-1.5}$ kpc using 3D-$N_{\rm H}$-tool\footnote{http://astro.uni-tuebingen.de/nh3d/nhtool} of \citet{doroshenko2024}. In any case, the source should be closer to us than 5 kpc.

From the non-detections in the VVV and \gaia data, We can put a constraint on the nature of the companion star. The extinction along the line of sight in the optical and infrared band is $A_V=6.36$ and $A_{\rm K}=0.61$, respectively \citep{doroshenko2024}. The \gaia G band and VVV Infrared survey have a detection limit of $m_{\rm G}=21$ and $m_{\rm Ks}=18.1$, respectively. Therefore, we obtained the absolute magnitude of the donor star should be $M_{\rm G}>2.3$ and $M_{\rm Ks}>5.1$. The donor star might be even fainter than the above estimates as often the optical emission from this type of system has a significant contribution from the accretion disk \citep{mukai2023}. Therefore, the upper limits of the optical and infrared magnitude suggest the companion star should be a low-mass main sequence star of spectra type later than A7/F0. Our \xmm observation has simultaneous UV observation with UVW1 (291 nm) and UVM2 (231 nm) filters. We did not find any counterparts in the \xmm UV data. 

\section{Discussion}
The source XMMU J173029.8--330920 was discovered in March 2023 during our \xmm campaign. Shortly afterward, in April 2023, we triggered our pre-allocated \nustar observation on following up on this source. We searched for a counterpart in \swift and \chandra source catalog but found no source association. We also looked for optical, infrared counterparts in \gaia and \emph{2MASS} catalogs, but no counterparts were found. 

When the X-ray spectra are fitted by a single Galactic absorption component, leaves residual in the ratio plot. The X-ray spectra are best fitted when we incorporate a partial covering component into the spectral models. The partial covering absorber has a column density of $N_{\rm H,pcf}=(1.5-2.3)\times10^{23}\ \rm cm^{-2}$. The requirement of a partial covering is commonly seen in the spectra of magnetic CVs \citep{ezuka1999,mondal2022,mondal2023}. A partial covering absorption suggests a fraction of the photons are seen directly and the rest through an intervening absorber, which suggests the absorber size is of the same order as the X-ray emitting region located closer to the source. A simple absorber with column density of $10^{23}\ \rm cm^{-2}$ is so high that it completely absorbs the photons below 2 keV, yet most IPs show a good level of emission in the soft X-ray band 0.2--2 keV. Therefore, \citet{norton1989} introduced the concept of a partial covering absorber that improved the spectral fit and explained the observed energy-dependent spin modulation. The total Hydrogen absorption column density ($\rm H_I$+$\rm H_2$)\footnote{https://www.swift.ac.uk/analysis/nhtot/} along the line of sight is $N_{\rm \sc H_I+H_2}=1.4\times10^{22}\ \rm cm^{-2}$ \citep{willingale2013}. The Galactic absorption towards the source obtained from X-ray spectral fitting is $(2-3)\times10^{22}\ \rm cm^{-2}$, which is slightly higher than the total Hydrogen absorption column density along the line of sight $N_{\rm \sc H_I+H_2}$. 

A pulsation period of $521.7\pm0.8$ s was found in both \xmm and \nustar data sets. The period is likely to be associated with the spin period of the WD. Such a spin period is typical for an IP. Typically, the spin period of IPs ranges from $\sim30$ s to $\sim3000$ s with a median value around $\sim1000$ s \citep{scaringi2010}. The high energy 5--10 keV pulsed fraction of the source is $42\pm16\%$. The PF in IPs decreases with increasing energy. This is due to the fact that the light curve of IPs shows increasing modulation depth with decreasing energy \citep{norton1989}, and photoelectric absorption has been considered in some part responsible for this trend. Most IPs do not show large PF at higher energies; however, the source GK Per \citep{norton1989}, AO Psc, and FO Aqr \citep{hellier1993} show PF of the order of 25--35\% at energies above 5 keV which is similar to the PF detected in 5--10 keV band for the source XMMU J173029.8–330920.

The source is most likely a strong candidate for IP. The upper limits estimate of the optical and infrared magnitude indicate the source is hosting a low-mass companion star. The optical extinction of $A_{V}=6.36$ along the line of sight is not high enough for non-detection of the optical/infrared counterpart if the source contains a high-mass companion. Therefore, ruling out the neutron star high-mass X-ray binary (NS HMXRB) scenario. Furthermore, the source spectrum is extremely soft with $\Gamma\sim1.9$, which is not typically seen in NS HMXRBs. The NS HMXRBs do not show strong ionized 6.7 and 6.9 keV lines and their spectra are best fitted by an absorbed power law model. On the other hand, the spectra of the source XMMU J173029.8–330920 are best fitted by an apec model that is typically used for fitting the spectra of IPs. Lastly, the source is also unlikely to be an ultra-compact X-ray binary (UCXB). The UCXBs are low-mass X-ray binaries usually containing a WD donor and a neutron star accretor with an ultra-short orbital period (<1 hr). The emission from NS-UCXBs is well characterized by blackbody emission with $kT\sim1-3$ keV originating from the NS surface \citep{koliopanos2015}. In our case, a blackbody component does not fit the \xmm+\nustar 0.5--50 keV spectrum and leaves excess above 10 keV in the ratio plot. Furthermore, a partial covering model is still required when fitting with the blackbody model, which is not typically seen when fitting the spectra of NS-UCXB. In addition to that the NS-UCXBs show little to no emission of Fe \citep{koliopanos2021}.

\subsection{WD Mass Measurement}
As J1730 is an IP whose X-ray emission primarily originates from an accretion column, we measure the WD mass fitting {\tt MCVSPEC} to the broadband X-ray spectra. First introduced in \citet{Vermette2023}, {\tt MCVSPEC} is an {\tt XSPEC} spectral model developed for magnetic CVs assuming a magnetically confined 1D accretion flow. Largely based on the methodology of \citet{saxton2005}, {\tt MCVSPEC} computes plasma temperature and density profiles varying along the accretion column between the WD surface and shock height. The model outputs an X-ray spectrum with thermal bremsstrahlung continuum and atomic lines by integrating X-ray emissivity over the accretion column. The input parameters for {\tt MCVSPEC} are $M$, $\dot{m}$ (specific accretion rate [g\,cm$^{-2}$\,s$^{-1}$]), $R_{m}/R$ where $R_{m}$ is the magnetospheric radius and $R$ is the WD radius, and $Z$ (abundance relative to the solar value).

For IPs where the accretion disk is truncated by WD magnetic fields, we assume that the free-falling gas gains kinetic energy from the magnetospheric radius ($R_m$) to the shock height ($h$). At the shock height, the infalling gas velocity reaches $v_{ff} = \sqrt{2GM \left(\frac{1}{R+h} - \frac{1}{R_m}\right)}$, which exceeds the sound speed and forms a stand-off shock. As the shock temperature is directly related to $M$ through $kT_{\rm shock} = \frac{3}{8} \mu m_{\rm H} v^2_{ff}$, both $R_m$ and $h$ affect the X-ray spectrum from the accretion column. In addition to {\tt MCVSPEC}, the {\tt reflect}, {\tt gaussian} and {\tt pcfabs} models in {\tt XSPEC} are incorporated to account for X-ray reflection by the WD surface and X-ray absorption by the accretion curtain. A Gaussian line component for the Fe K-$\alpha$ fluorescent line is frozen at 6.4 keV with a fixed width of $\sigma = 0.01$ keV. For the {\tt reflect} model, we linked the reflection scaling factor ($r_{refl}$) to the shock height $h$, which was determined self-consistently by the {\tt MCVSPEC} model in each spectral fit, following the recipe suggested by \citet{Tsujimoto2018}. Note that X-ray reflection is more pronounced when the accretion column is shorter. Overall, the full spectral model was set to {\tt tbabs*pcfabs*(reflect*MCVSPEC + gauss)}. 

Before fitting the X-ray spectra, we constrained the range of input parameters, such as $R_m/R$ and $\dot{m}$. Following \citet{shaw2020}, we constrained $R_m/R$ by assuming that the WD is in spin equilibrium with the accretion disk at the magnetospheric radius. In this case, the magnetospheric radius is given by $R_m = {\left(\frac{GMP^2}{4 \pi^2}\right)}^{1/3}$, where $P = 521$ s is the WD spin period. The specific accretion rate, however, is largely unconstrained due to the unknown source distance ($d$) and fractional accretion column area ($f$). The specific accretion rate is defined as $\dot{m} = \frac{\dot{M}}{4\pi f R^2}$ where $\dot{M}$ is the total mass accretion rate [g\,s$^{-1}$]. Assuming that X-ray emission represents most of the radiation from the IP (i.e., $L \approx L_X$), $\dot{M}$ can be calculated from $L = GM\dot{M} \left(\frac{1}{R} - \frac{1}{R_{m}}\right)$, where $L = 4\pi d^2 F_X$ is the total luminosity and $F_X$ represents the unabsorbed X-ray flux. We assumed $d = 1.4\rm{-}5$ kpc which is estimated from the Galactic absorption column density towards the line of sight from X-ray spectral fitting. We first estimated the minimum specific accretion rate $\dot{m}_{\rm min}$ by assuming the shortest source distance (1.4 kpc) and maximum fractional accretion column area ($f_{\rm max}$). $f_{\rm max} = \frac{R}{2R_m}$ is given for a dipole B-field geometry, assuming that the infalling gas originates from the entire accretion disk truncated at $R_m$ \citep{2002apa..book.....F}.

Since some of the parameters described above depend on the WD mass, we adopted a handful of $M$ values, namely, 0.6, 0.8, 0.9, 1.0, 1.1, 1.2, 1.3, and $1.35 M_\odot$ as initial estimates. For each WD mass estimate, we computed $R_m/R$ and the $\dot{m}_{\rm min}$. We then proceeded to fit the X-ray spectra by inputting these $R_m/R$ and the $\dot{m}_{\rm min}$ values. This is not sufficient since the specific accretion rate may be larger than $\dot{m}_{\rm min}$ due to potentially larger source distance ($d \simgt 1.4$ kpc), smaller fractional accretion column area ($f \simlt f_{\rm max}$) and missing source luminosity below the X-ray band ($L \simgt L_X$). Therefore, we also considered a large range of $\dot{m}$ above $\dot{m}_{\rm min}$ ($\dot{m} = 0.0043\rm{-}83$ [g\,cm$^{-2}$\,s$^{-1}$]). For each $\dot{m}$ value (frozen), we fit the X-ray spectra and determined the WD mass and statistical errors. We found that the fitted WD mass becomes saturated and independent of $\dot{m}$ at higher $\dot{m}$ values (typically $\simgt 10$ [g\,cm$^{-2}$\,s$^{-1}$]) when the shock height becomes only a small fraction of the WD radius. This saturation of the WD mass over $\dot{m}$ results from the fact that both the shock temperature and X-ray reflection (both of which are functions of $h/R$) become less dependent on the shock height. 

\begin{figure}
    \centering
    \includegraphics[width=\figsize\textwidth]{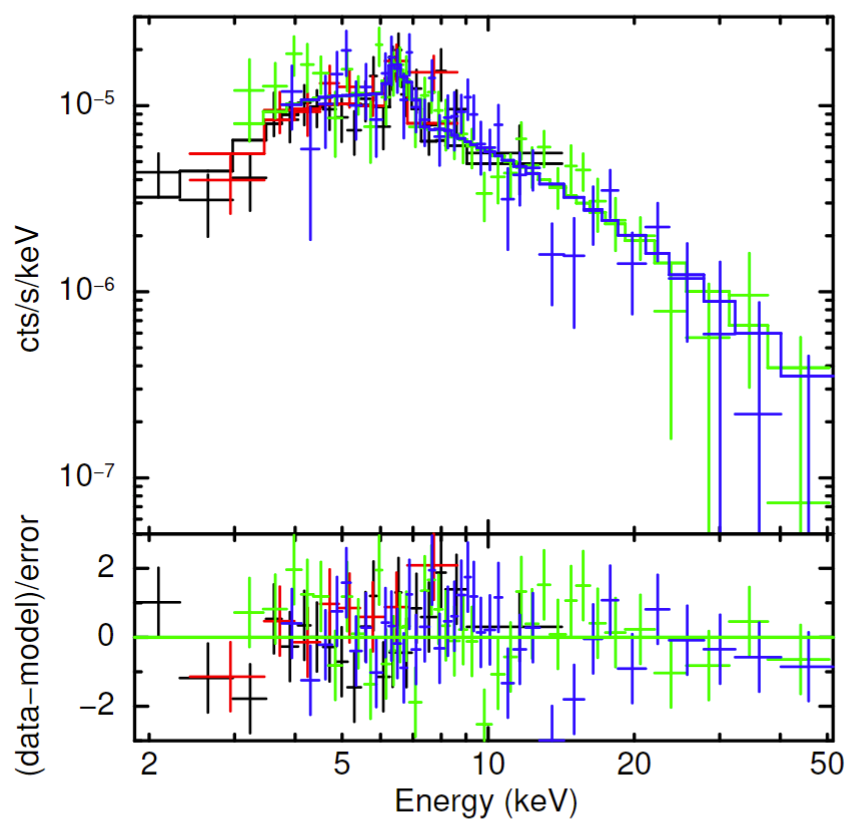}
    \caption{The \xmm\ and \nustar\ spectra of XMMU J173029.8--330920, fit with {\tt tbabs*pcfabs*(reflect*MCVSPEC + gauss)}. This fit utilized an initial mass of $1.2 M_\odot$ and thus an $R_m/R$ ratio of $26$. The specific accretion rate for this fitting was 15 [g\,cm$^{-2}$\,s$^{-1}$], resulting in the best-fit WD mass of $\sim 1.28^{+0.12}_{-0.23} M_\odot$, which is consistent with the initially assumed value of $1.2\, M_\odot$. The black, red, green, and blue data points are from EPIC-pn, MOS2, FPMA, and FPMB, respectively.}
    \label{fig:spect}
\end{figure}

In each case, our model was well fit to the X-ray spectra with reduced $\chi^2 = 1.18\rm{-}1.21$ (Fig. \ref{fig:spect}). In the end, we checked the self-consistency of our fitting results by ensuring the measured WD mass was consistent with the assumed WD mass within statistical and systematic errors. The same method was applied to fit broad-band X-ray spectra of another IP obtained by \xmm\ and \nustar\ (Salcedo et al. submitted to ApJ). The systematic errors stem from the range of the specific accretion rate we considered. 
We found that the initial mass estimates of $M \ge 1.0 M_\odot$ yielded self-consistent WD mass measurements. We were able to constrain the WD mass range to $0.94\rm{-}1.4\,  M_{\odot}$, which is comparable to or larger than the mean WD mass of magnetic CVs \citep[$0.81^{+0.16}_{-0.20} M_\odot$;][]{pala2022}. The large WD mass range is due to statistical and systematic errors, largely associated with the unknown source distance and fractional accretion column area. As emphasized in \citet{Vermette2023}, it is desirable to constrain the source distance well and obtain better photon statistics (particularly above 10 keV) for determining the WD mass more accurately in the future.

\subsection{Fe-K-$\alpha$ lines}
There is an indication of Fe $K_{\alpha}$ line emission at 6.4 keV. The equivalent width of the line is $312\pm104$ eV. Most IP-type sources also display the presence of ionized Fe lines at 6.7 (Fe XXV) and 6.9 (Fe XXVI) keV with an equivalent width higher than 50 eV \citep{xu2016}. We checked the presence of the 6.7 and 6.9 keV lines and estimated an upper limit on the equivalent width. The upper limit on the equivalent width of Fe XXV and Fe XXVI lines are 134 and 147 eV, respectively, which can be confirmed with a longer observation in the future. The 6.4 keV iron $K_{\alpha}$ line is mainly created by irradiation of neutral material by hard X-ray photons. Compact objects accreting matter through the accretion disk show X-ray reflection from the accretion disk. Whereas in the case of accreting WD, most of the hard X-rays are produced so close to the compact object that one might expect half of the photons to be directed toward the WD surface and reflected back to the observer. X-ray reflection in accreting compact objects shows the signature of Fe $K_{\alpha}$ fluorescence line and a Compton reflection hump in 10--30 keV. The Compton reflection hump was originally difficult to measure; however, with the launch of the \nustar satellite, the X-ray reflection was detected in a couple of magnetic CVs \citep{mukai2015}. We also checked the presence of a reflection component in the X-ray spectrum of XMMU J173029.8--33092 by adding a convolution term \texttt{reflect} \citep{magdziarz1995} to the total model. However, the addition of the reflection component does not provide any improvement in $\Delta\chi^2$. On the other hand, the neutral Fe $K_{\alpha}$ line could have been produced by reprocessing when the hard X-ray passes through the partial covering medium. The hard X-rays traveling through the absorbing material will interact with the iron and create a Fe $K_{\alpha}$ fluorescence line. In such a scenario, the intensity of Fe $K_{\alpha}$ line will increase with the thickness of the ambient material. In a sample study of mCVs using \suzaku data, the column density of the partial absorber seems to correlate with the 6.4 keV line intensity \citep{eze2014,eze2015}, which indicates that the emission of the neutral iron $K_{\alpha}$ is primarily caused by absorption-induced fluorescence from the absorber.

\section{Conclusions}
We serendipitously discovered the source XMMU J173029.8--330920 during our \xmm Heritage survey of the Galactic disk. Later, we followed up on the sources using our pre-approved \nustar ToO observation. We suggest that the identity of XMMU J173029.8--330920 is an IP based on \xmm and \nustar data analysis. The X-ray spectra can be fitted by emission from collisionally ionized diffuse gas with plasma temperature of $26^{+11}_{-5}$ keV. One requires a partial covering absorption in addition to the Galactic absorption to model the X-ray spectrum. The X-ray spectra show the presence of neutral Fe $K_{\alpha}$ emission line at 6.4 keV, while the X-ray light curves show periodic modulation with a period of $521.7\pm0.8$ s. These features are commonly seen in accreting magnetic CVs of IP type.


\begin{acknowledgements}
SM, and GP acknowledge financial support from the European Research Council (ERC) under the European Union’s Horizon 2020 research and innovation program HotMilk (grant agreement No. 865637). SM and GP also acknowledge support from Bando per il Finanziamento della Ricerca Fondamentale 2022 dell’Istituto Nazionale di Astrofisica (INAF): GO Large program and from the Framework per l’Attrazione e il Rafforzamento delle Eccellenze (FARE) per la ricerca in Italia (R20L5S39T9). KM is partially supported by NASA ADAP program (NNH22ZDA001N-ADAP). 

\end{acknowledgements}

%
%

\bibliographystyle{aa} 
\bibliography{refs}

\end{document}